# Refining Platelet Purification Methods: Enhancing Proteomics for Clinical Applications

Vibecke Markhus [1,2,*], Katarina Fritz-Wallace[2], Olav Mjaavatten[2], Einar K. Kristoffersen[3], Dorota Goplen[1], Frode Selheim[2]

**Affiliation**

1. Department of Oncology, Haukeland University Hospital, Bergen, Norway.
2. Proteomics Unit of University of Bergen (PROBE), University of Bergen, Jonas Lies vei 91, 5009 Bergen, Norway.
3. Department of Immunology and Transfusion Medicine, Haukeland University Hospital, Bergen, Norway
4. * Correspondence: Vibecke.markhus@uib.no

## Abstract

*Background*: Platelet proteomics has significant potential in clinical research, but the lack of an ideal method for isolating high-purity platelets remains a challenge. Efficient platelet isolation is crucial for accurate proteomic analysis, yet current methods often result in contamination or platelet loss, affecting data reliability.

*Objectives:* Our study aimed to optimize a platelet isolation technique that allows for the preparation of uncontaminated platelet samples with minimal loss. Additionally, we sought to determine the most effective mass spectrometry-based proteomic method for analyzing platelet proteins in terms of coverage and sensitivity.

*Methods:* We refined a platelet isolation protocol by evaluating centrifugation time to minimize the required blood volume while preserving high platelet purity and yield. With this refined method, we assessed three mass spectrometry-based proteomic approaches:
- Label-free Quantification with Data-Independent Acquisition (LFQ-DIA)
- Label-free Quantification with Data-Dependent Acquisition (LFQ-DDA)
- Chemical Labeling with Tandem Mass Tag (TMT-DDA)

*Results:* Among the tested methods, LFQ-DIA provided the most comprehensive and sensitive analysis of platelet proteins compared to LFQ-DDA and TMT-DDA. Additionally, our findings revealed significant age-related differences in platelet protein composition, underscoring the importance of using age-matched control groups in large studies investigating platelet characteristics and disease-related biomarkers.

***Conclusions:*** Our optimized platelet isolation technique provides a cost-effective and efficient method for preparing high-purity platelet samples for proteomic analysis. Furthermore, LFQ-DIA emerges as the most suitable mass spectrometry approach for platelet proteomics. The observed age-related platelet protein differences highlight the necessity of age-matching in platelet-based biomarker studies.

**Keywords:** Platelet Proteomics. Platelet Isolation. Mass Spectrometry. Label-free Quantification. Data-Independent Acquisition (DIA). Data-Dependent Acquisition (DDA). Tandem Mass Tag (TMT).

**Introduction**

Platelets play a critical role in hemostasis and are involved in a wide range of physiological and pathological processes, including cancer, inflammation, atherosclerosis, and angiogenesis. Their dynamic and adaptable proteome not only supports their essential functions in normal physiological conditions but also enables them to participate in disease progression. In particular, during the development of solid tumors, alterations in the platelet proteome can reflect changes in the tumor microenvironment and may serve as valuable biomarkers for early disease detection, prognosis, and monitoring therapeutic responses. This makes platelet proteomics a promising area of research, offering potential insights into the molecular mechanisms driving disease and advancing clinical diagnostic capabilities [1,2,3,4].

To investigate these changes, obtaining high-purity platelet samples free from red blood cell and leukocyte contamination is essential. This study presents an optimized isolation protocol that reduces contamination, shortens processing time, and enhances practicality for clinical applications. Our approach provides high-purity, unactivated platelets that are well-suited for proteomic analysis [2,5,9-11].

Liquid Chromatography-Mass Spectrometry (LC-MS) is an effective tool for detecting protein variations related to diseases, though standardized protocols for platelet proteomics have yet to be fully established. We applied Gu-HCL (guanidine hydrochloride) and chemical labeling across all three data-independent acquisition methods to enhance protein extraction, identification, and quantification, improving sensitivity and accuracy.

## Materials and Methods

### Sample collection and anticoagulation

Human whole blood samples were donated from 10 healthy volunteers. Age spread from 37 to 74. The study is approved by the Regional Committee for Medical and Health Research Ethics (Nr 457396) in Norway. All patients were thoroughly informed, and each patient signed a written consent. The study was conducted in accordance with the Declaration of Helsinki.

Blood samples were collected in tubes (G455055 Vacuette®ACD-A9 mL from Med-Kjemi AS, Norway) containing ACD solution (12mM citric acid, 15mM sodium citrate, 25mM D-glucose). A total of 8 ml in each tube (7 ml whole blood and 1 ml ACD solution). Apyrase (0.01 U/mL) and EGTA (3mM) were added to the final concentrations to prevent platelet activation. (All chemicals were from Sigma-Aldrich.com, USA unless otherwise specified).

### Platelet isolation and purification

Platelet isolation and purification were performed as previously reported, with certain modifications [5]. The whole blood was centrifuged at 330 g for 10 minutes at room temperature (RT). The components were separated into three layers: platelet poor plasma (PPP), platelet rich plasma (PRP), and red blood cells (Figure 1). The PRP fraction was carefully extracted to avoid contamination from the other layers.

The PRP was diluted 1:1 with 1xPBS (Phosphate Buffered Saline, pH 7,4). This solution was centrifuged at 240 g for 10 minutes (RT). The supernatant was centrifuged at 430 g for 10 minutes (RT). The centrifugation times were carefully chosen to minimize contamination from white and red blood cells, thereby ensuring a higher purity of platelets for subsequent analysis.

The platelet pellets were suspended in CGS buffer (120mM sodium chloride, 12.9 mM trisodium citrate, 30mM D- glucose, pH 6.5) and centrifuged for 10 minutes at 430 g (RT). The platelet pellets were washed twice with CGS to remove all remains of plasma proteins.

The platelets, red blood cells and white blood cells were counted using an automated cell counter Sysmex XN-9100 at the Laboratory Clinic at Haukeland Hospital, Bergen following the manufacturer's instructions.

The platelet pellets were stored at - 80°C for subsequent proteomics analysis.

**Protein identification by LC-MS/MS** *(this procedure applies for 20 µg proteins)*

Isolated platelet pellets were resuspended in 150 µl hot Guanidine-HCL (Gu-HCl), and vortexed (RPM 400) at 95 °C for 10 min. Samples were sonicated 3 times for 30 sec with 30 sec breaks in between and then centrifuged in accuSpinMicro 17R from Fischer Scientific® at 4 ºC at 16 200 g for 10 min.
The supernatant was transferred to 1.5 ml Eppendorf Protein Lo-Bind® tubes and the protein concentration was determined with BCA Protein Assay Kit (Pierce®, Thermo Scientific®) according to the manufacturer's instructions.

We took a volume equal to 20 µg protein, added 5 µl 50 mM Tris (2-carboxyethyl) phosphine (TCEP), 5 µl 0.1 M 2-Chloroacetamide (CAA) and then adjusted the volume with 6 M Gu-HCl to a total of 50 µl. This solution was boiled for 10 min at 95 ºC (RPM 400) to ensure reduction and alkylation of cysteine residues. Proteins digested with Lys-C (Enzyme: protein ratio 1:100) and incubated at 37 ˙C for 1 h (RPM 400). Then digested with trypsin (Enzyme: protein ratio 1:50) and incubated at 37 ˙C for 16 h (RPM 400)

The solution was acidified with 4 µl 10 % Trifluoroacetic acid (TFA) and added 0.1 % TFA to dilute sample to 300 µl. Peptides were then desalted using Oasis HLB 96-well µElution Plate, 2 mg Sorbent per Well, 30 µm particle size (Thermo Fischer Scientific®), and the desalted sample frozen at -80 °C before lyophilized (freeze dryer).

**LC-MS/MS analysis**

Before LC-MS/MS acquisition the samples were split into 3 parts: one aliquot for TMT labeling and the other for LFQ-DDA and LFQ-DIA analysis.

For TMT labelling, the tryptic peptides were dissolved in 100 mM HEPES (4-(2-hydroxyethyl)-1-piperazineethanesulfonic acid) and labelled with the TMT 10plex kit (Thermo Fisher Scientific®) for 1h at room temperature. The labeling reaction was quenched by the addition of hydroxylamine to a final concentration of 0.4 %. Labelled samples were mixed, and the mix was fractionated into 6 fractions using Pierce High pH Reversed-Phase Peptide Fractionation Kit (Thermo Fisher Scientific®).

Samples were dissolved in 20 µl 0.5 % Formic Acid (FA), 2 % Acetonitrile (ACN), and peptide concentration was measured with a NanoDropOne Spectrometer (Thermo Scientific®). Less than 1 µg (0.5 µg for DDA, 0.8 µg for DIA) was injected and separated on the LC and detected either using the data-dependent acquisition (LFQ-DDA and TMT-DDA), or data-independent acquisition (LFQ-DIA). Detailed information of peptide analyzed by LC-MS/MS is described in Supplementary Files S1. Raw data has been deposited in the ProteomeXchange consortium with the dataset identifier PXD058404.

**Data Analysis**

The Orbitrap Eclipse™ Tribrid™ Mass Spectrometer (Thermo Fisher Scientific®) was used in this study. Details are given in the supplemental (Supplementary Files S1).

DIA data signal extraction and quantification were performed using Spectronaut $^{TM}$ (Biognosys®, v18.6) with standard default settings. DDA and TMT analysis used Proteome Discover Program $^{TM}$ (Thermo Fisher Scientific®, v. 2.5) with standard settings. Details are given in the supplemental Methods.

Further statistical analysis of DIA, DDA and TMT data was performed by Perseus software (version 2.0.11). The normalized protein abundances were transformed to a logarithmic scale ($\log2(x)$) and the filter for valid values was set to 70%. False discover rate (FDR) of < 1%.

Categorical annotation was used to set up a group with young vs old test subjects. Median age young group: 53 years (Range from 37 to 59 years). Median age-old group: 70 years (Range from 67 to 74 years). We used a two sample Welch T-test ($p<0,01$) to find significant different fold change for the group. Hierarchical clustering was performed with the Euclidean function and complete linkage in Perseus.

The identified protein names were compared and illustrated using Venn diagram (Venny 2.1.0 csic.es, Figure 2-A). To explore the impacts of regulated proteins and to visualize their interactions we used Funrich [12] [Figure 2-B, 2-C, 2-D].

## Results and Discussion

*Platelet Purity and Contamination Control*

Platelets possess a distinct proteomic profile, and contamination from white blood cells (WBCs) and red blood cells (RBCs) can lead to erroneous disease biomarker identification [1, 4]. For high-precision platelet proteomics, contamination should be minimized, ideally below 1% for RBCs and even lower for WBCs [1, 4, 5].

To assess the impact of centrifugation time on plasma and platelet recovery, as well as contamination levels, we analyzed samples following the centrifugation step at 330g. The results were as follows: 5 minutes: 57 x [$10^9$/L] platelets (SD ±19, n=4) and 0.8 mL of plasma. 7 minutes: 76 x [$10^9$/L] platelets (SD ±26, n=6) and 1.0 mL of plasma. 10 minutes: 89 x [$10^9$/L] platelets (SD ±21, n=6) and 1.3 mL of plasma.

WBC and RBC contamination was detected in only one of seven experiments at 5 and 7 minutes (WBC: 0.01%, RBC: 1% of platelet count), while both were below detection limits after 10 min centrifugation. Scatter plot analysis confirmed an unactivated platelet population devoid of WBC and RBC contamination (Supplementary File 2).

Our rapid differential centrifugation protocol achieved at least 99.99% WBC purity. This result aligns with previous studies utilizing gradient centrifugation and filtration techniques [9, 20 - 21]. Our approach significantly reduces processing time while maintaining high platelet purity, making it ideal for clinical applications. While cost-effective and scalable, differential centrifugation carries a risk of contamination if not carefully executed. Alternative purification methods include: Immuno-affinity-based protocols. This gives high specificity but is expensive and complex. Microfluidic techniques. This is efficient but unsuitable for large-volume samples.

*Comparative Analysis of Platelet Proteomics Methods*

Mass spectrometry (MS)-based proteomics has advanced significantly with improvements in liquid chromatography (LC) and MS instrumentation. To determine the most effective platelet proteomic approach, we assessed three different data-independent methodologies: LFQ-DIA, LFQ-DDA, and TMT-DDA [12]. LFQ-DIA identified 4,743 proteins, including 728 unique proteins. TMT-DDA identified 4,232 proteins, with 451 unique proteins. LFQ-DDA identified 3,663 proteins, with 225 unique proteins. Common proteins across all three methods: 3,092 All data illustrated in Figure 2 and Supplementary Tables S1.

The choice of DDA, DIA, or TMT for clinical research depends on factors such as reproducibility, throughput, quantification accuracy, and cost. In general DIA (Data-Independent Acquisition) is a good choice for clinical proteomics because of its high reproducibility, low missing values, and ability to handle large cohorts. TMT (Tandem Mass Tag) is useful for controlling clinical studies where high-throughput multiplexing is required, but ratio compression can limit accuracy. DDA (Data-Dependent Acquisition) is less ideal for clinical applications due to its stochastic nature and missing data issues (Figure 4) [27,28].

*Age-Related Differences in Platelet Proteomes*

We investigated age-related differences in platelet protein expression using a Welch T-test ($p<0.01$). We identified 53 differentially expressed proteins, with 36 enriched in younger donors (mean age: 53 years) and 17 in older donors (mean age: 70 years). Key findings include Methylated-DNA-protein-cysteine-methyltransferase (MGMT): Two-fold higher in older donors, with promoter methylation linked to various cancers [13, 14]. Paraoxonase 3 (PON3): Three-fold higher in younger donors, known for inhibiting LDL oxidation and atherosclerosis progression [15]. Fibulin 1 (FBLN1): Two-fold higher in younger donors, associated with tumor suppression, hemostasis, and thrombosis [16] (Figure 3, Supplementary Table S1).

*Implications for Disease Research and Personalized Medicine*

Platelets play a central role in diseases such as cancer, cardiovascular disorders, and inflammation. Proteomics facilitates the identification of disease-specific proteins, aiding both diagnostics and therapeutic strategies. However, contamination from WBCs, RBCs, or plasma proteins can skew results, necessitating rigorous purification [9, 17-21].

Age-related proteomic differences highlight the importance of accounting for biological variables in biomarker research. Platelet reactivity increases with age, elevating thrombotic risk in older individuals, while gender-specific variations add further complexity. Recognizing these factors enhances research validity and supports personalized treatment strategies [23, 25-26].

Understanding age-related proteomic changes is critical for advancing disease diagnostics and therapeutic interventions, emphasizing the necessity of well-matched control groups in clinical studies.

**Conclusion:**

Future studies should continue to explore the implications of age differences in platelet proteomics, particularly in larger, more diverse cohorts. Our finding highlights the critical need for age-matched control groups in studies exploring platelet characteristics or platelet-derived biomarkers of disease. Additionally, the integration of LFQ-DIA with other omics approaches, such as genomics and metabolomics, could provide a more comprehensive understanding of the role of platelets in disease, as well as the discovery of novel biomarkers for early diagnosis, prognosis, and therapeutic monitoring.

**Declaration of interest**

The authors have no relevant affiliations or financial involvement with any organization or entity with a financial interest in or financial conflict with the subject matter or materials discussed in the manuscript. This includes employment, consultancies, honoraria, stock ownership or options, expert testimony, grants or patents received or pending, or royalties.

## Acknowledgments

We acknowledge Helse-Vest for financial support. We are also grateful to the volunteers who participated in this study. LC-MS/MS was performed at the Proteomics Unit at the University of Bergen. PROBE is a member of the National Network of Advanced Proteomics Infrastructure (NAPI), which is funded by the Research Council of Norway (INFRASTRUKTUR-program project number: 295910).

# Figures/Tables:

**Figure 1:** Schematic overview of the isolation of platelets process outlined in this protocol

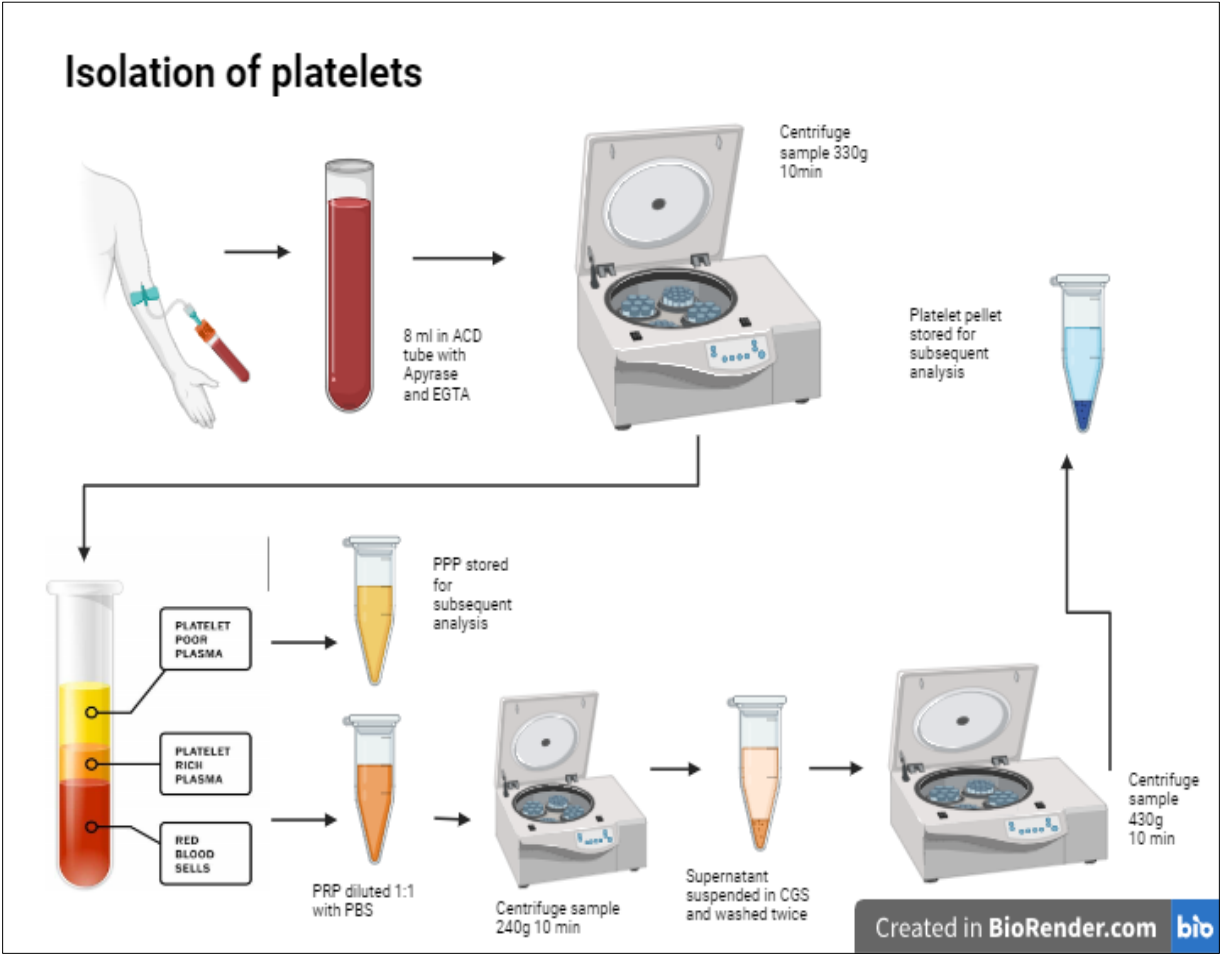

**Figure 2: MS-based methods for platelet proteome analysis.** (A) Venn diagram for identified platelets proteins for LFQ-DIA, LFQ-DDA and TMT-DDA. Numbers of proteins detected in at least seven out of ten healthy donors are shown: LFQ-DIA (4743), LFQ-DDA (3663), TMT-DDA (4232).

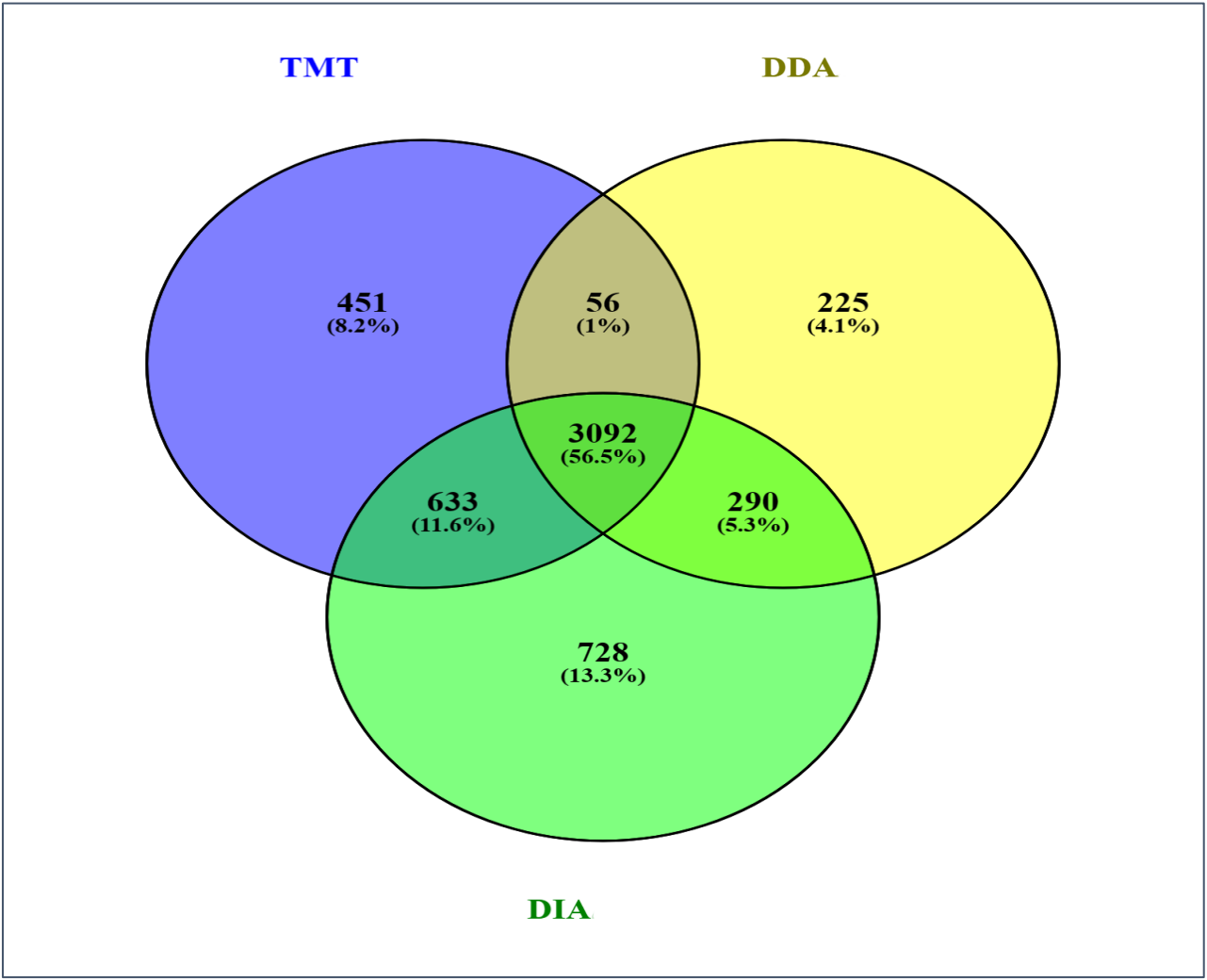

**Figure 3: Age-related protein expression in human platelets.** (A) Heatmap of 53 significantly differentially expressed age-related proteins (two sample Welch T-test (p<0.01). We introduce three interesting proteins. (B) High abundance of the Methylated-DNA-protein-cysteine- Methyltransferase protein (MGMT) in old healthy donors (mean 70 year). (C) Low abundance of the Paraoxonase 3 protein (PON3) in old healthy donors (mean 70 years). (D) High abundance of Fibulin 1 protein (FBLN 1) in young healthy donors (mean 53 years

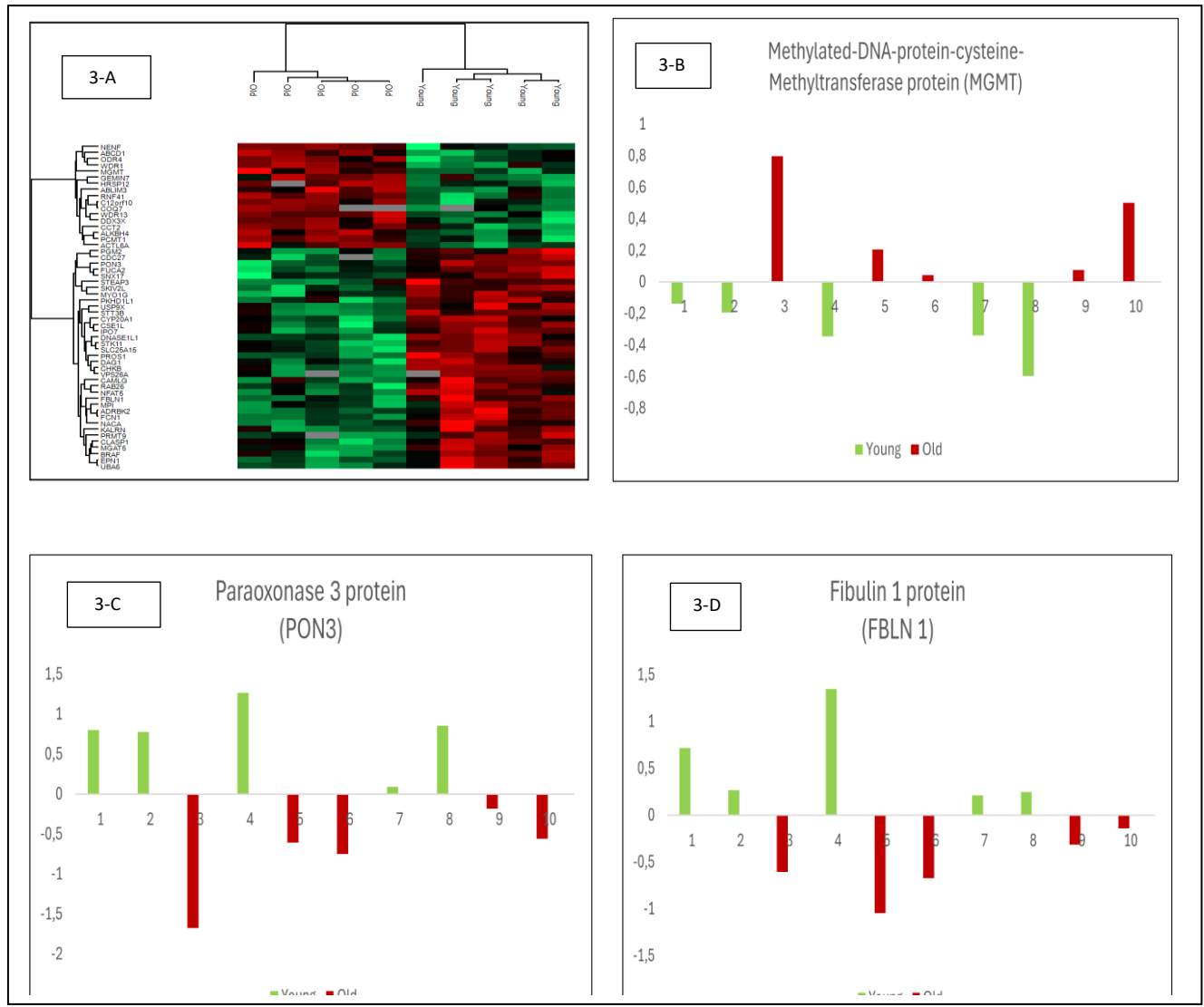

**Figure 4: Comparison for Clinical Research**

| Feature | DDA | DIA | TMT |
|---|---|---|---|
| **Reproducibility** | ❌ Low (stochastic selection) | ✅ High (systematic selection) | ✅ High (multiplexing) |
| **Quantification Accuracy** | ✅ High (for selected peptides) | ✅ High (consistent peptide detection) | ⚠️ Medium (ratio compression issue) |
| **Throughput** | ⚠️ Moderate | ✅ High | ✅ Very High |
| **Missing Data** | ❌ Common | ✅ Minimal | ✅ Minimal |
| **Best Use Case** | Small-scale discovery | Large-scale proteomics, biomarker validation | High-throughput studies, but less for absolute quantification |

---

Why DIA is Preferred for Clinical Proteomics?

✅ Reproducibility – Consistent peptide quantification across multiple clinical samples.
✅ Minimal Missing Data – Unlike DDA, DIA captures all ions systematically.
✅ Scalability – Suitable for large patient cohorts.
✅ Data Retrospectability – Raw data can be re-analyzed later using updated libraries.

# Peptide analysis of DDA, DIA and TMT (SUPPLEMENTARY FILES S1)

## NanoLC-Orbitrap Eclipse mass spectrometry at PROBE

Peptides were dissolved in 2% acetonitrile (ACN), 0.5% formic acid (FA), were injected into an Ultimate 3000 RSLC system (Thermo Scientific, Sunnyvale, California, USA) connected online to an Orbitrap Eclipse mass spectrometer (Thermo Scientific, Bremen, Germany) equipped with EASY-spray nano-electrospray ion source (Thermo Scientific).

## Trapping and desalting

The sample was loaded and desalted on a pre-column (Acclaim PepMap 100, 2cm x 75µm ID nanoViper column, packed with 3µm C18 beads) at a flow rate of 5µl/min for 5 min with 0.1% trifluoroacetic acid.

## Eclipse-DDA/DIA LFQ LC RUN (195 min)

Peptides (0.8µg) were separated during a biphasic ACN gradient from two nanoflow UPLC pumps (flow rate of 200 nl/min) on a 50 cm analytical column (PepMap RSLC, 50cm x 75 µm ID. EASY-spray column, packed with 2µm C18 beads). Solvent A and B were 0.1% FA (vol/vol) in water and 100% ACN respectively. The gradient composition was 5%B during trapping (5min) followed by 5-7%B over 1min, 7–22%B for the next 119min, 22–32%B over 27min, and 32–85%B over 3min. Elution of very hydrophobic peptides and conditioning of the column were performed for 15 minutes isocratic elution with 85%B and 20 minutes isocratic conditioning with 5%B respectively. Instrument control was through Thermo Scientific SII for Xcalibur 1.6.

## Eclipse-DDA TMT LC RUN (105 min), 2-method wash

Peptides were separated during a biphasic ACN gradient from two nanoflow UPLC pumps (flow rate of 250 nl/min) on a 25 cm analytical column (PepMap RSLC, 25cm x 75 µm ID. EASY-spray column, packed with 2µm C18 beads). Solvent A and B were 0.1% FA (vol/vol) in water and 100% ACN respectively. The gradient composition was 5%B during trapping (5min) followed by 5-7%B over 1 min, 7–22%B for the next 74min, 22–30%B over 10min, and 30–85%B over 2min. To minimize peptide carry-over, injection valve was switched to loading position after 3min (valve: 6_1) and back to injection position after 4min (valve: 1_2) while pumping at 85% B, followed by a 15min conditioning program injecting 6 µl 2-propanol with valve in the injection position to further clean the system before the next injection (valve switched to load position after 5 minutes to condition the trap column). Instrument control was through Thermo Scientific SII for Xcalibur 1.6.

### High field asymmetric waveform ion mobility spectrometry (FAIMS)

The FAIMS filter performs gas-phase fractionation, enabling preferred accumulation of multiply charged ions to maximize acquisition efficiency. FAIMS results in cleaner MS2 spectra and better signal-to-noise ratios. Short-ion residence time in the FAIMS Pro interface electrode assembly enables use of multiple CV settings in a single run to increase proteome coverage.

### Ionsource parameter

The spray and ion-source parameters were as follows: ion spray voltage = 2000V, no sheath and auxiliary gas flow, and capillary temperature = 275 °C,

### DIA with FAIMS

Peptides eluted from the column were detected in the Orbitrap Eclipse Mass Spectrometer with FAIMS enabled using two compensation voltages (CVs) of -45V and -65V. The mass spectrometer was operated in the DIA-mode (data-independent-acquisition) to automatically switch between one full scan MS and MS/MS acquisition of 60 mass segments with a cycle time of 3s. Instrument control was through Orbitrap Eclipse Tune 3.5 and Xcalibur 4.5. MS spectra were acquired in the scan range 375-1400-1000 m/z with resolution R = 120 000 at m/z 200, automatic gain control (AGC) target of 4e5 and a maximum injection time (IT) set to 50ms. After 20 windows an additional MS scan was performed. Using an isolation window of 10 Da (first window 400.43-410.44 m/z and the last from 990.7-1000.7 m/z), all ions in the m/z window were sequentially isolated to a target value (AGC) of 4e5 and a maximum IT of 54 ms in the C-trap before HCD fragmentation (Higher-Energy Collision Dissociation). Fragmentation was performed with a normalized collision energy (NCE) of 30 %, and fragments were detected in the Orbitrap at a resolution of 30 000 at m/z 200, with scan range fixed at m/z 145-1450 m/z. Lock-mass internal calibration was not enabled.

### DDA with FAIMS

Peptides eluted from the column were detected in the Orbitrap Eclipse Mass Spectrometer with FAIMS enabled using three compensation voltages (CVs), -45V, -65V, and -80. During each CV, the mass spectrometer was operated in the DDA-mode (data-dependent-acquisition) to automatically switch between one full scan MS and MS/MS acquisition. Instrument control was through Orbitrap Eclipse Tune 3.5 and Xcalibur 4.5. The cycle time was maintained at 0.9s, 1.2s, and 0.9s/CV respectively. MS spectra were acquired in the scan range 375-1500 m/z with resolution R = 120 000 at m/z 200, automatic gain control (AGC) target of 4e5 and a maximum injection time (IT) set to Auto. The most intense eluting peptides with charge states 2 to 5 were sequentially isolated to a target value (AGC) of 5e4 and a maximum IT of 75 ms in the C-trap, and isolation width maintained at 1.6 m/z (quadrupole isolation), before fragmentation in the HCD (Higher-Energy Collision Dissociation. Fragmentation was performed with a normalized collision energy (NCE) of 30 %, and fragments were detected in the Orbitrap at a resolution of 15 000 at m/z 200, with the first mass fixed at m/z 110. One MS/MS spectrum of a precursor mass was allowed before dynamic exclusion for 30s with "exclude isotopes" on. Lock-mass internal calibration was not enabled.

## TMT – MS3 with Real-Time Search

Peptides eluted from the column were detected in the Orbitrap Eclipse Mass Spectrometer with FAIMS enabled using two compensation voltages (CVs), -50V and -70. During each CV, the mass spectrometer was operated in the DDA-mode (data-dependent-acquisition) to automatically switch between one full scan MS and MS/MS acquisition, MS2 spectra were searched in real time against the human swissprot database containing 20423 entries, downloaded in April 2023. Only identified MS2 spectra were then used in MS3 acquisition for TMT quantitation. Instrument control was through Orbitrap Eclipse Tune 3.5 and Xcalibur 4.5. Detailed instrument settings are described in table S1.

| Acquisition Settings | DDA – MS2 | DDA – MS3 RTS | DIA |
|---|---|---|---|
| CVs | -45, -65, -80 | -50, -70 | -45, -65 |
| Top speed | 0.9, 1.2, 0.9 | 3, 3 | 1.5, 1.5 |
| RF lens | 30 | 30 | 30 |
| Orbitrap MS1 resolution | 120k | 120k | 120k |
| Scan range (m/z) | 275-1500 | 400-1600 | 375-1400 |
| Standardized MS1 AGC target | 100% | 100% | 100% |
| MS1 max IT (mode) | Auto | Auto | Auto |
| Charge state | 2-5 | 2-6 | 2-5 |
| Dynamic exclusion | 30s | 45s | 30s |
| MS2 resolution | 15k | Turbo | 30k |
| MS2 Scan range | First mass 110 | 400-1600 | 400-1000 |
| MS2 Isolation Window | 1.6 | 0.7 | 10 |
| Standardized MS2 AGC target | 100% | 100% | 1000% |
| MS2 max IT (mode) | 75ms | 35ms | 54ms |
| MS2 HCD/CID NCE% | HCD 30% | CID 30% | HCD 30% |
| SPS MS3 resolution | - | 50k | - |
| MS3 Scan range | - | 100-500 | - |
| SPS MS3 isolation window | - | 0.7 | - |
| Standardized SPS MS3 AGC target | - | 200% | - |
| SPS MS3 max IT (mode) | - | 200ms | - |
| SPS MS3 HCD NCE% | - | 55 | - |
| SPS MS3 Notches | - | 10 | - |
| DIA window size | - | - | 10 |
| DIA window number | - | - | 60 |

## Supplementary Section S2: Scatterplot

The scatter plot revealed an unactivated platelet population with no detectable traces of WBC or RBC. Y-axis shows forward scatter (FSC) and X-axis represents side fluorescence (SFL).

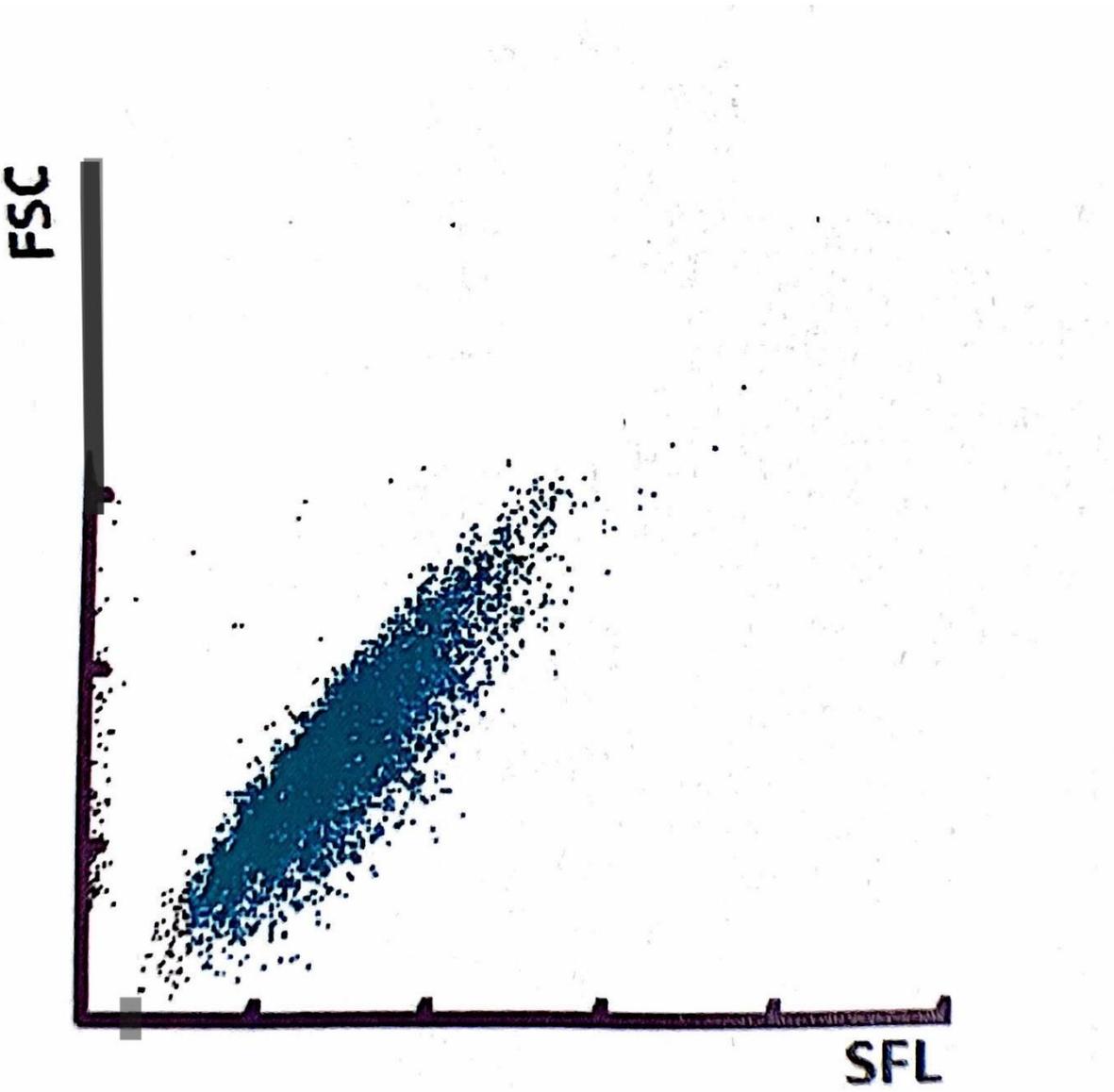